\documentclass{article}

\usepackage{cleveref}
\usepackage{amsmath}
\usepackage{graphicx}
\usepackage{comment}
\usepackage{subfloat}
\usepackage{subfig}
\usepackage{geometry}
\usepackage{tabularx}
\usepackage{listings}
\usepackage{sidenotes}
\usepackage{amssymb}
\usepackage[figuresright]{rotating}

\begin{document}

\title{Microgrid - The microthreaded many-core architecture}

\author{Irfan Uddin\\
University of Amsterdam, The Netherlands\\
mirfanud@uva.nl}

\maketitle

\begin{abstract}

Traditional processors use the \emph{von Neumann} execution model, some other
processors in the past have used the \emph{dataflow} execution model. A
combination of \emph{von Neuman} model and \emph{dataflow} model is also tried
in the past and the resultant model is referred as \emph{hybrid dataflow}
execution model. We describe a hybrid dataflow model known as the
microthreading. It provides constructs for creation, synchronization and
communication between threads in an intermediate language. The microthreading
model is an abstract programming and machine model for many-core architecture.
A particular instance of this model is named as the microthreaded architecture
or the Microgrid. This architecture implements all the concurrency constructs
of the microthreading model in the hardware with the management of these
constructs in the hardware.

\end{abstract}

\setcounter{tocdepth}{1}
\tableofcontents

\newpage

\section{Introduction}
\label{sn:introduction}

Traditional processors are based on the \emph{von Neumann} execution model. In
this model a sequence of instructions is executed one-by-one and the state of
the program is identified by a single program counter. However, performance can
not be improved by executing instructions sequentially. In order to improve
the performance, \emph{dataflow} scheduling is used where instructions are
executed based on the availability of data. Theoretically, dataflow models are
parallel execution models, because instructions can be scheduled only with the
availability of the data, and the scheduled instructions can be executed
independently. Most of the out-of-order execution techniques are derived from
dataflow scheduling. However, the execution of the program is still determined
largely by the instruction sequence as dataflow scheduling is only applied to
a few instructions in a small window over the sequential code. In contrast,
Moonsoon~\cite{Papadopoulos:1998:MET:285930.285999} and
Wavescalar~\cite{Swanson:2007:WA:1233307.1233308} are based on dataflow
models.

The \emph{von Neumann} model is simple but sequential, the \emph{dataflow}
model is parallel and improves the efficiency of the execution of the program
but with the cost of adding complexity in the hardware design, therefore a
\emph{hybrid dataflow} model~\cite{5009501,Iannucci:1988:TDN:633625.52416}
was tried to combine the advantages of both. The issue with generalized
dataflow model is the requirement of a matching store to detect which
operations become ready to execute when its dependencies are satisfied. The
efficient organization of the matching store is not so clear to the designers
of the execution models. With a hybrid dataflow model, the ordering of
instructions can be sequentialized to reduce the need for a big matching store.
In hybrid dataflow models instructions are executed based on the \emph{von
Neumann} model within a thread but there is support for dynamic execution of
threads based on the \emph{dataflow} model. This model allowed multicore
architectures to exploit parallelism in programs.
P-RISC~\cite{Razdan:1994:PSA:645462.655024},
Multiscalar~\cite{Sohi:1995:MP:225830.224451} and
DDM-CMP~\cite{stavrou:ddm-cmp:} are based on the \emph{hybrid dataflow}
model.

Modern many-core systems one way or other provide concurrency constructs to
exploit parallelism. To quote Prof. Chris Jesshope \emph{By 2020 we could see
up to $10^4$ cores and $10^6$ hardware threads on a single chip}. The
concurrency constructs introduced by the microthreading model are implemented
in the instruction set of the microthreaded architecture which is also referred
as the Microgrid. This architecture assumes a lot of concurrency in the
applications and provides the concurrent architecture to exploit parallelism.

The rest of the paper is organized as follows. In Section~\ref{sn:latency_tolerance}
we define the latency tolerance feature in computer architecture. We explain
the details of the microthreading model in Section~\ref{sn:microthreading_model} and
the details of the architecture based on this model in Section~\ref{sn:microgrid}. We
describe the programming of the Microgrid in Section~\ref{sn:program_microgrids}. We
explain the microthreading model in the context of the Microgrid
in Section~\ref{sn:microthreading_microgrid}. 
We present the I/O in the Microgrid in Section~\ref{sn:io_microgrid} and conclude the
paper in Section~\ref{sn:conclusion}.

\section{Latency tolerance}\label{sn:latency_tolerance}

\begin{figure}
\begin{centering}
    \includegraphics[width=0.7\textwidth]{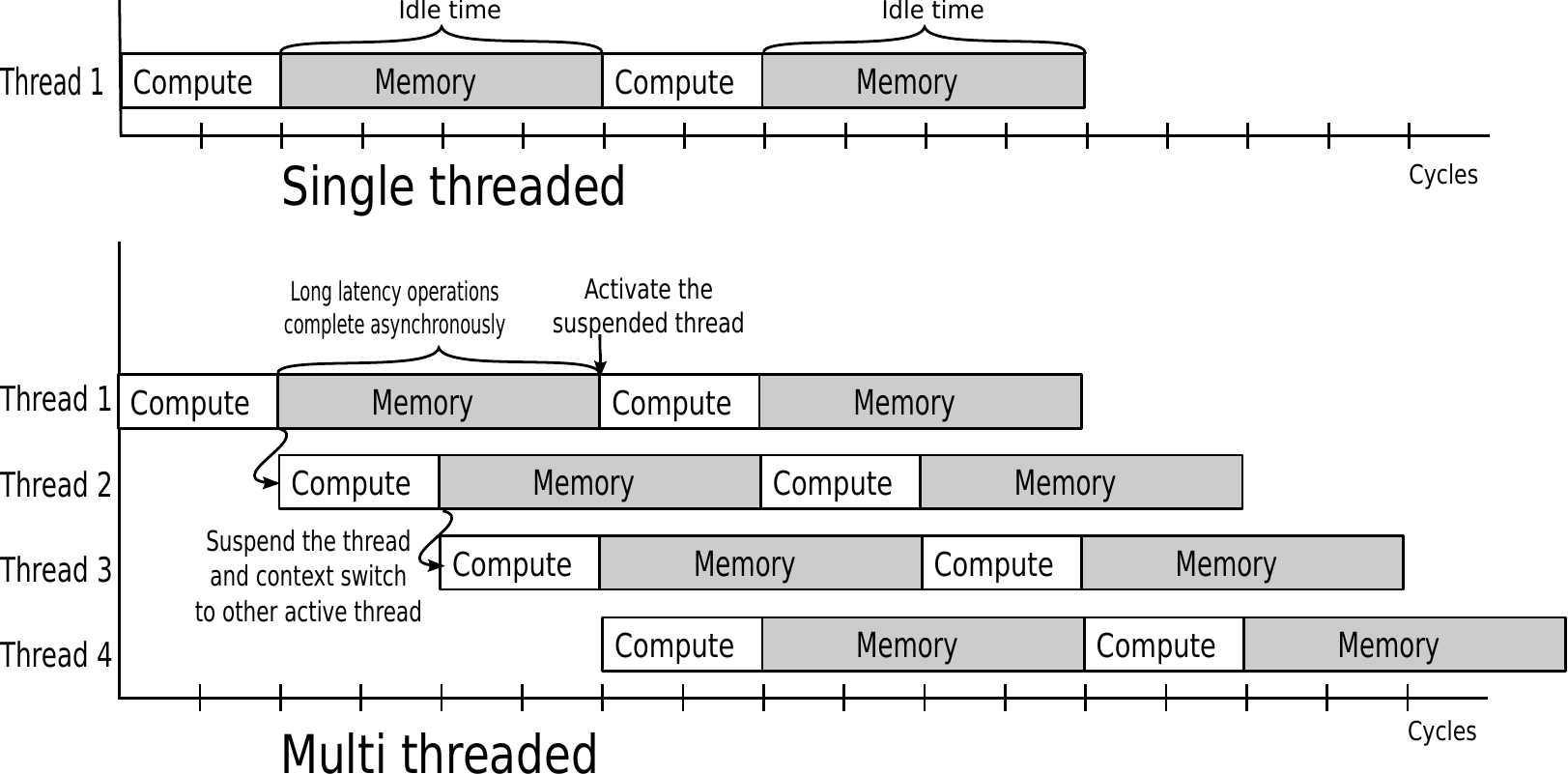}
    \caption{\label{fig:latency_tolerance}Latency tolerance in multi-threaded programs assuming multiple
    instructions issued and executed in the core.}
\end{centering}
\end{figure}

In any program, a computation is preceded and followed by memory operation
which takes a variable amount of time as it depends on the locality of the data
i.e. the data is located in L1-, L2-,L3- caches or off-chip memory. In
single-threaded programs the processor has to wait for memory operations to
complete and then continue with the computation. In multi-threaded programs,
when a memory operation is issued, the thread may be suspended and execution is
switched to another available thread. Because of dataflow scheduling, the
memory operation completes asynchronously and wakes up the suspended thread.
The execution of program is tolerant to the long latency operations and is
shown in fig.~\ref{fig:latency_tolerance} assuming multiple instructions issued
and executed in a single core.

The use of multiple threads per core in order to hide latency has been
understood for long time~\cite{Weber:1989:EBM:74926.74956}. For instance,
software time sharing is adequate to tolerate external I/O latencies but can
not tolerate other long latency operations (e.g. memory access) because of the
absence of fine-grained latency tolerance. Hardware multi-threading with
dynamic scheduling is used to tolerate long latency operations e.g. MTA cores
can switch on every miss operation but it requires a long pipeline flush.
Niagara cores switch on the issue of a memory operation~\cite{5434030} and
therefore tolerate long latency operations, but can not tolerate the latency in
FPUs or other asynchronous operations (e.g. management of threads).

\section{The microthreading model}\label{sn:microthreading_model}

The microthreading model~\cite{Jesshope:2000:MNA:784563.784568} is based on a
type of \emph{hybrid dataflow} model and has evolved from
DRISC~\cite{Bolychevsky96dynamicscheduling} (Dynamically-scheduled RISC) which
was proposed in 1996 with the goal of separating computation from
communication. DRISC provides dataflow scheduling in RISC core, which executes
instructions asynchronously and with multiple threads it can tolerate the
latency of long latency operations. The innovation in this model is that
instead of multiple threads, an ordered set of threads referred as families are
used, which provide composability in programs. The operations to create
families resemble the fork/join operations found in most parallel programming
models. The microthreading model is actually a hybrid of von Neumann model,
dataflow scheduling of threads. 

\begin{figure}
\begin{centering}
    \includegraphics[width=0.5\textwidth]{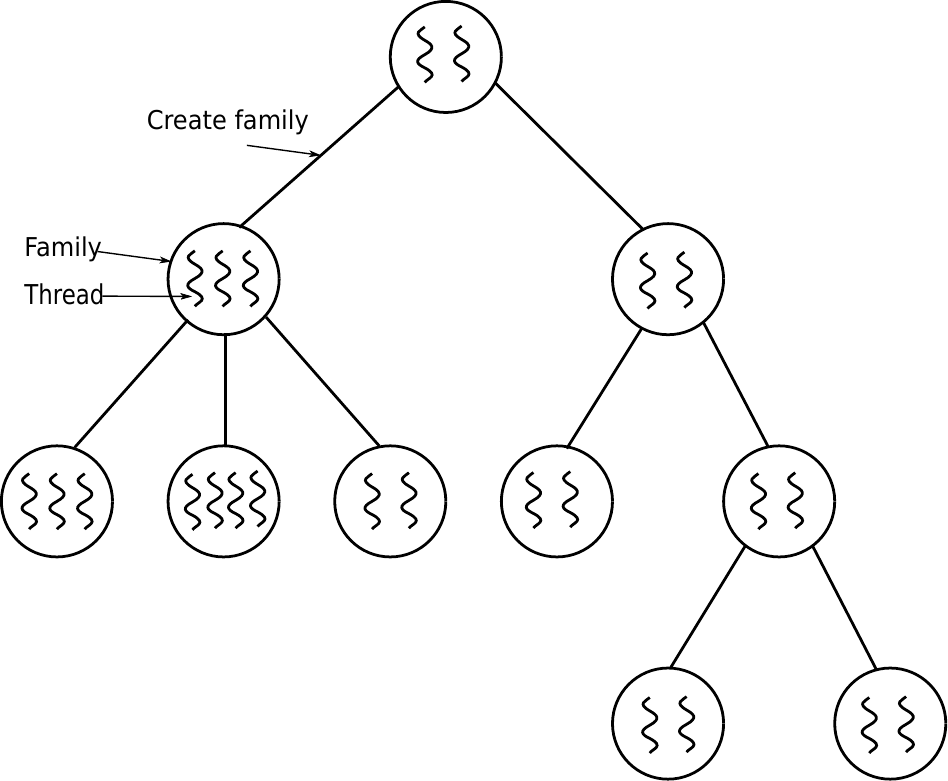}
    \caption{\label{fig:svp_concurrency}An example concurrency tree created by the microthreading model. A family
    can create any combination of threads and the hierarchy of families can go
    until any level.}
\end{centering}
\end{figure}

The microthreading model has been refined over the decade from a single
processor model to an abstract machine model for many-cores and can capture as
much concurrency as possible using families of threads in a dynamically
evolving concurrency tree. A family is comparable to loop or function call in
traditional programming. Any thread can create further families (any
heterogeneous combination is supported by the model) showing a hierarchy of
threads. An example concurrency tree of the microthreading model is shown
in fig.~\ref{fig:svp_concurrency}, where every family is composed of some number of
threads and every thread can create another family.

As the overhead of creating and synchronizing threads in software is 10 to 100
thousands cycles on contemporary hardware~\cite{T3-2011}. The microthreading
model shifts the perspective from software threads to hardware threads with
concurrency management in hardware in order to reduce the cost from few
thousands cycles to just a few cycles.  The number of cycles taken by an
instruction depends on the dynamic state of the architecture, but because of
asynchronous completion and fine-grained latency tolerance the model has the
potential to achieve the goal of RISC i.e. \emph{one cycle per instruction}. In
the microthreading model the throughput of the program can potentially
demonstrate that every instruction takes one cycle to complete, assuming
single instruction issued and executed in the core.

\subsection{Communication and synchronization}
\label{sn:communitation_synchronization}

The microthreading model supports hybrid dataflow scheduling by using
I-structures~\cite{Arvind:1989:IDS:69558.69562}. An I-structure is a data
structure with the semantics of dataflow i.e. every element has a state of
either \emph{full} or \emph{empty}. Any operation accessing an element with
\emph{empty} state is suspended. The state of the element is changed to
\emph{full} asynchronously, and the suspended operation is released to access
the element. This process is termed as \emph{split-phase}
operation~\cite{Lin98thedesign} which has two phases; requesting and consuming.
A request to the desired data is issued, but if the data is not available, the
request is suspended. The execution model can continue executing other
computations while the request is in the progress. When the data arrives, the
suspended operation is released and the instruction can consume the data. 

\begin{figure}

\begin{centering}

\includegraphics[width=0.7\textwidth]{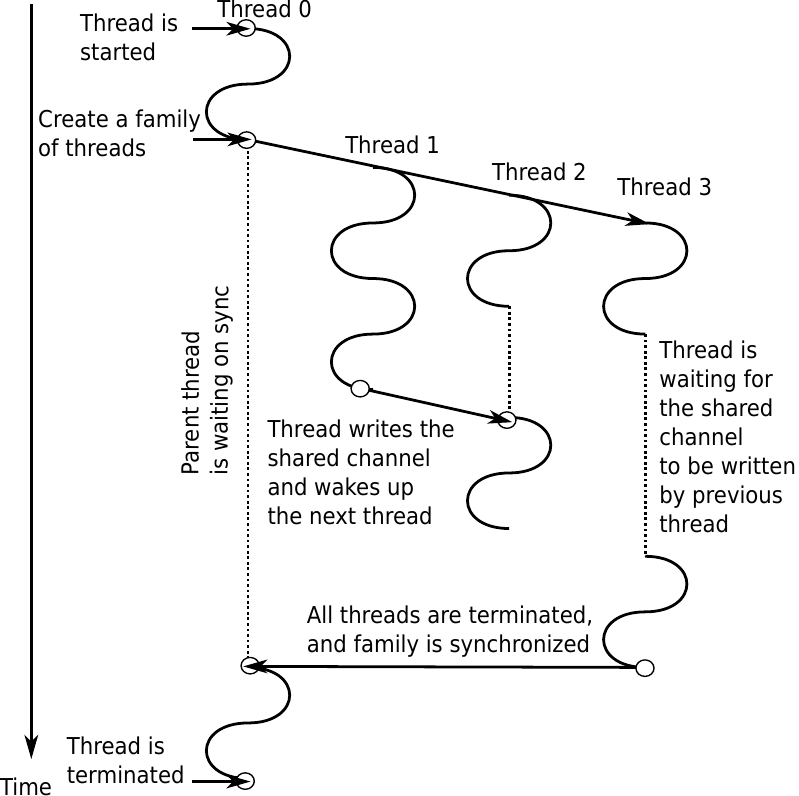}

\caption{\label{fig:com-sync}The communication between threads and the
synchronization of the family.}

\end{centering}

\end{figure}

The microthreading model uses I-structure as a set of channels; globals and
shared to support communication between threads. These channels are actually
registers address using the register names of the underlying ISA. But for the
sake of generality these registers are referred as channels. These channels
have blocking read and non-blocking writes. By blocking we mean that the
operation is suspended because of the unavailability of the data, and by
non-blocking we mean that the operation is performed and the writing of the
data complete asynchronously. Global channels are mapped to all threads in the
family and have read only access. Shared channels provide the mechanism for the
communication only between adjacent threads uni-directionally in the ordered
sequence of threads in the family. The sharing of synchronizers between threads
enables fast thread-to-thread synchronization, for example to implement
dependencies in a loop. The creating thread communicate with the created
threads using messages, which can access the synchronizing memory of a family
remotely. The one-way communication is restricted but ensures a deadlock free
communication~\cite{Vu:2007:FSV:1775348.1775375}. The communication and
synchronization between threads are shown in fig.~\ref{fig:com-sync} and more
details can be found in~\cite[Sec.~4.3.3.3]{poss.12}. 

\subsection{Memory consistency model}

The microthreading model addresses consistency using two models for threads and
families. A single thread is sequentially consistent, as all the instructions
within a thread appear to execute in sequential order. A family is weakly
consistent, as it follows the three properties of Weak Consistency:

\begin{itemize}

\item Any created thread can perform a read or write operation only after all
    the writes by the creating thread prior to the creation of the family.

\item The family can be synchronized only when all previous write operations by
    the created threads are performed.

\item The creating thread can perform a read or write operation only after the
    created family is synchronized.

\end{itemize}

Memory consistency models~\cite{Gharachorloo:1990:MCE:325096.325102,
Mosberger:1993:MCM:160551.160553} have been defined for different programming
and machine models. We present only two consistency models which are relevant
to the microthreading model.

\begin{itemize}

\item \textbf{Sequential Consistency (SC)} :

\noindent A system is sequentially consistent if the result of the execution of
instructions is the same as if instruction of all the cores were executed in
some sequential order. In addition, the instructions of each individual core
appear in the same sequence as specified by the program. SC has two cases:

\begin{enumerate}

\item A read operation by any processor/thread is allowed to perform only if
    all previous read and write accesses are performed globally.

\item A write operation by any processor/thread is allowed to perform only if
    all previous read and write accesses are performed globally.

\end{enumerate}

\item \textbf{Weak Consistency (WC)} :

\noindent A system is weakly consistent if the synchronizing operation is
performed only if there is no pending ordinary (non-synchronizing) read or
write operations, and any ordinary read or write operations can not be
performed if there is a pending synchronizing operation. The weak consistency
is defined in~\cite{Gharachorloo:1990:MCE:325096.325102} but we are slightly
modifying the terminology to match with the microthreading model. WC has three
cases:

\begin{enumerate}

\item All previous synchronizing operations must be performed before an
    ordinary operation is performed.

\item All previous ordinary operations must be performed before the
    synchronizing variable.

\item Synchronizing operations are sequentially consistent with respect to one
    another.

\end{enumerate}

\end{itemize}

\subsection{The notion of resources in the microthreading model}

The microthreading model is an abstract machine and programming model and
therefore does not have physical resources, but it is designed with the notion
of resources as the model addresses many-core architectures. A family of
threads is allocated to a group of resources and are referred as
\emph{place}. These resources can be one or more cores, single or many threads
etc. A family created on a \emph{place} will execute there until synchronized
and a thread allocated to a core will execute there until terminated. 

In order to control the number of threads on a core per family,
\emph{windowsize} is introduced as a run-time parameter. A programmer can
carefully use this parameter to use only the requested number of resources,
leaving resources for other families (or may be more important families). More
importantly this parameter is used to avoid deadlock from dependencies down the
concurrency tree. It can wisely be used by the programmer or compiler to ensure
that at least one leaf of the concurrency tree can continue execution by using
the available resources and hence avoid deadlocks. The \emph{windowsize} limits
the number of threads executing in a family, similar to the concept as
k-bounded loops~\cite{Culler:1988:RRD:633625.52417} in dataflow scheduling. The
parameter 'k' is comparable to the \emph{windowsize} in the microthreading
model. However, it should be noted that \emph{windowsize} is per core and 'k'
is per loop which may be distributed on many cores.

The microthreading model introduces \emph{break} used by a thread to terminate
the execution of the family. The \emph{break} will let the already created
threads of the family to complete and will stop the creation of any new threads
in the family. It is comparable to the \emph{exit} statement in loops or
functions in traditional programming, and is useful when the synchronization of
the family depends on some dynamic conditions.

\subsection{Concurrency constructs}\label{sn:conc_constructs}

The microthreading model defines concurrency constructs to use the concurrency
provided by the hardware. But these constructs need to be introduced in some
intermediate language in order to write programs. In this section we explain
some of the concurrency constructs of the microthreading model and show that
these constructs are defined in an intermediate language referred as
SL~\cite{poss.12, poss.12.sl}. In Section~\ref{sn:program_microgrids} we will show an
example program that will demonstrate the use of concurrency constructs of the
microthreading model in an application. 

\subsection*{Allocation and deallocation of place}

A group of cores i.e. \emph{place} is required to be allocated before the
creation of a family and released when the family is synchronized.

\begin{itemize}

\item $sep\_alloc(<parameters>)$ and $sep\_free(<parameters>)$ \\ are used to
    allocate and de-allocate cores through a software API. The $<parameters>$
    are a list of parameters passed for allocation and de-allocation e.g.
    specifying the strategy of allocation or number of cores etc. and can be
    found in~\cite{SEP}.

\end{itemize}

\subsection*{Creation and synchronization of family}
\label{sn:creation_synchronization}

A family of threads can be created by using $sl\_create$ constructs similar to
a loop in traditional programming. The created family can be synchronized using
$sl\_sync$ construct. 

\begin{itemize}

\item $sl\_create(fid, pid, start, limit, step, windowsize, options, thread [,
    <arguments>])$ \\defines the creation of a parameterized family.

\item $fid$ \\ is the identifier of the family of type $sl\_family\_t$.
    
\item $pid$ \\ is the identifier of the \emph{place} of type $sl\_place\_t$.

\item $start$, $limit$ and $step$ \\ indicate the starting, ending and
    iteration step counter of threads (similar to the iterators in a loop).

\item $windowsize$ \\ determines the upper limit on the number of threads that
    can be created on a core.

\item $option$ \\ decides the way the family should be created an executed.
    e.g. $sl\_forceseq$ will force the family to execute sequentially.

\item $thread$ \\ is the name of the thread which defines the code to be
    executed by created threads in the family.

\item $<argument>$ \\ is a comma separated list of global and shared arguments
    passed to the thread.

\item $sl\_sync()$ \\ is used for the synchronization of the family.

\end{itemize}

\subsection*{Global and shared channels}

The global and shared channels introduced by the microthreading model are
implemented as global and shared variable in the intermediate language. They
are used as parameters/arguments to the creation of a family of thread.

\begin{itemize}

\item $sl\_glarg(type, variable, value)$ and $sl\_sharg(type, variable,
    value)$\\ indicates the global and shared $variable$ of $type$ with some
    $value$ passed from the creating thread to the created thread.

\item $sl\_glparm(type, variable)$ and $sl\_shparm(type, variable)$\\ defines
    the global and shared $variable$ of $type$ received by the created thread
    from the creating thread.

\item $sl\_getp(variable)$ \\ is used to read from global or shared
    $variable$.

\item $sl\_setp(variable, value)$ \\ is used to write the shared $variable$
    with a modified $value$ to be read by the next thread.

\item $sl\_seta(variable, value)$ \\ is used by the creating thread to write a
    $value$ to the shared variable to the first thread of the family.

\item $sl\_geta(variable)$ \\ is used by the creating thread to read the shared
    $variable$ from the last thread in the family. 

\end{itemize}

\subsection*{Starting and terminating threads}

Threads are implicitly created with the creation of a parameterized family and
terminated when all of their instructions are executed. But programmers need to
define the starting and terminating of a thread similar to the way functions
are defined in traditional programming.

\begin{itemize}

\item $sl\_def(type, <parameter>)$\\ is used to define a thread with return of
    $type$ and $<parameter>$ of a list of the global or shared variables used
    by the thread. 

\item $sl\_enddef$ \\ defines the terminating of the thread. 

\item $sl\_index(index)$ \\ can be used to retrieve the $index$ of the ordered
    set of threads in the family.

\end{itemize}

\subsection*{Breaking family}

Any thread in a family can use the break statement to stop creation of new threads
in the family. The syntax is given below:.

\begin{itemize}
\item $sl\_break$ \\is used to stop creating new threads in the family.
\end{itemize}

\subsection{More about families}

The microthreading model has evolved over the years and have introduced
different types of families as per the requirements of the programs. All the
families are created using $sl\_create$ construct (except detached family, see
below). The family can be of any of the type (described in this section)
based on the way the created threads are executing. However, some types of
families are need to be explicitly defined by passing a parameter to the create
construct. In this section we give a brief overview of the different types of
families supported by the microthreading model with the aim that programmers
can write a parallel program using different combination of threads and
families. Families that are required to be explicitly defined at the time of
creation are explicitly stated in their respective subsection.

\subsubsection*{Independent family}

A family is called independent family when its created threads do not require
any communication between threads. Embarrassingly parallel applications
generally do not require communication between created threads and the
microthreading model supports these types of applications by creating them as
independent families.

\subsubsection*{Dependent family}

Threads in dependent family communicate with each other. Fine-grained parallel
applications require a lot of communication and coarse-grained parallel
applications communicate a little between threads. The microthreading model
supports these types of application by creating dependent families. These
families communicate with each other through shared channels introduced earlier
(c.f. Section~\ref{sn:communitation_synchronization}). Dependent families are
inherently sequential and therefore do not get any speedup by distributing the
threads on many cores. But in the microthreading model these threads get
benefit from latency tolerance and asynchronous completion as a dependent
family can be executed along with other families. 

\subsubsection*{Homogeneous family}

All families are statically homogeneous but can be made dynamically
heterogeneous if the programmer calls/creates different functions in a thread
dependent on the index of the thread.

\subsubsection*{Heterogeneous family}

A family can be composed of threads that are not identical to each other. The
programmer need to explicitly write the code for different threads based on the
index of the thread to perform different operations. 

\subsubsection*{Detached family}

A family is required to be synchronized before continuing further. But there
may be situations when a family may not want to synchronize and is called
\emph{detached} family. For instance, creating a family that prints some
characters on terminal may not require synchronization as all the created
threads perform only read operations and do not modify any memory. The
programmer has to explicitly specify the \emph{detached} family by creating a
family with $sl\_detach(<parameters>)$ instead of $sl\_create(<parameters>)$,
where $parameters$ are defined in Section~\ref{sn:creation_synchronization}.

\subsubsection*{Exclusive family}

To support mutual exclusion in the microthreading model, exclusive families are
introduced. Exclusive families are treated differently to regular families
because of the requirement of being mutually exclusive. More details about
mutual exclusion in the microthreading model can be found
in~\cite[Sec.~14.1]{poss.12}. To create an exclusive family the programmer have
to explicitly use $sl\_exclusive$ in the place of $options$ in the create construct.
Generally exclusive families consist of a single thread.

\subsubsection*{Sequentialized family}

A family can be forced to execute sequentially, to avoid using resources by
creating many threads. The programmer has to explicitly use $sl\_forceseq$
at the place of $options$ in the create construct. The sequentialized family will
execute as regular code of the parent thread.

\section{The Microgrid} \label{sn:microgrid}

Microgrid is actually a blueprint with parameters defined for the architecture.
There are many Microgrids; with 1 core, 128 cores, 1000 cores, random bank
memory, COMA etc. We use the term Microgrid to generalize all instances of the
blueprint. At some time we talk about a particular instance, but then we
explicitly give the details of the parameters. In the context of the Microgrid,
a core is explicitly termed as microthreaded core to differentiate from
traditional cores. In addition, a thread is termed as microthread to
distinguish it from traditional threads. 

The basic component of the Microgrid is the microthreaded core, and it is
important that we understand the execution of instructions in the core in order
to simulate the core at the high-level. The concurrency constructs of the
microthreading model are implemented in the instruction set (ISA) of the core.
The cores are designed to be simple; single issue, in-order, 6-stage
pipeline~\cite{Jesshope03} and based on RISC. Some components of the operating
system are implemented in the silicon of the
core~\cite{Cotofana:1998:DCI:945405.938236} e.g. scheduler, resource
allocation, mapping etc. Some energy inefficient features are removed from the
design of the core e.g. speculation, out-of-order execution and cache
prefetching. 

\begin{figure}

\begin{centering}

\includegraphics[width=0.8\textwidth]{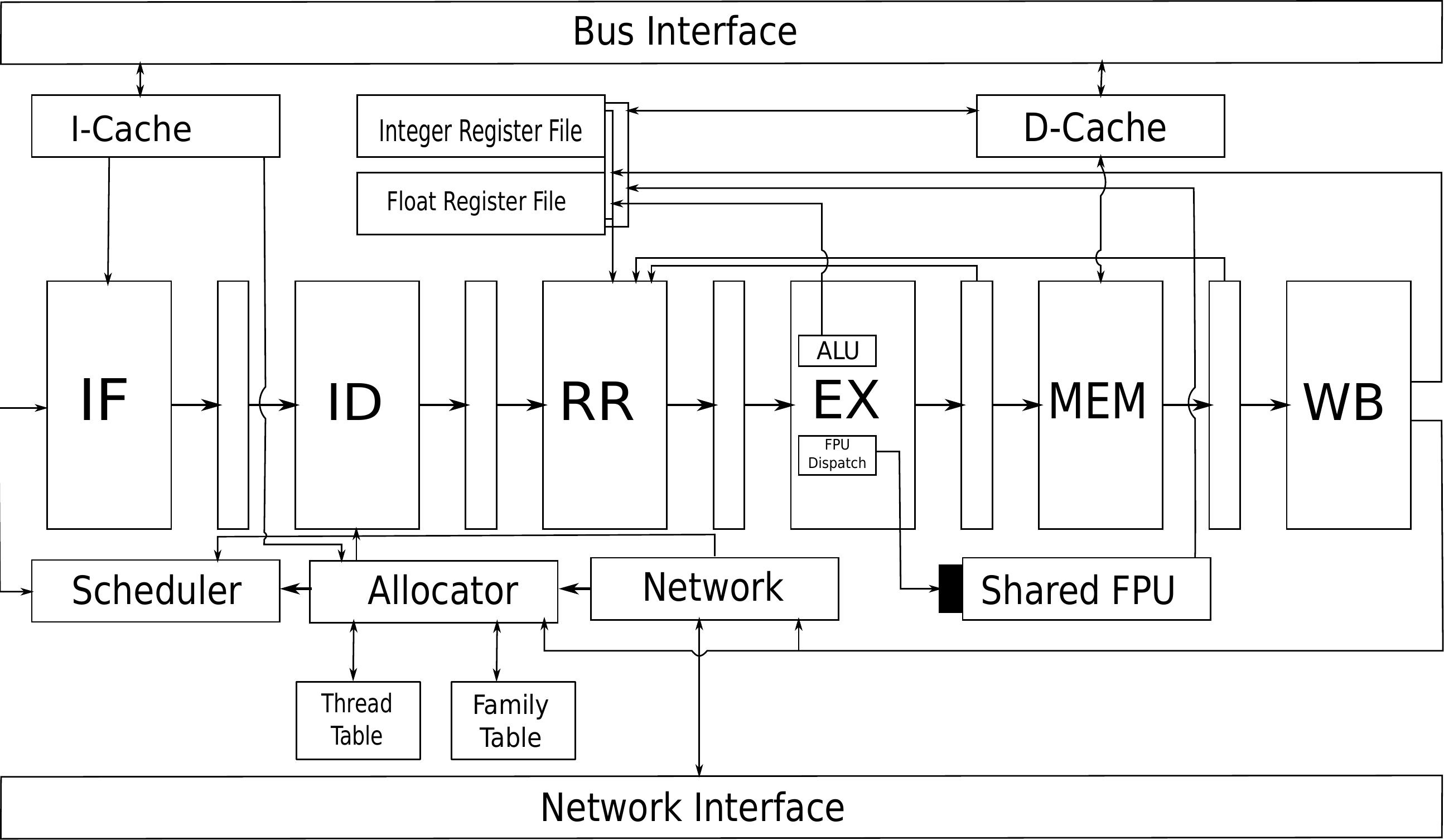}
 
\caption{\label{fig:mtcore}The microthreaded core.}

\end{centering}

\end{figure}

The 6 stage pipeline of the core is shown in fig.~\ref{fig:mtcore}, where we are
not showing the scale of components, but mainly the layout of components. It is
a classic RISC core with some modifications to existing components and adding
some more components to support the concurrency constructs of the
microthreading model in the core. The instruction fetch (IF), instruction
decode (ID), register read (RR), execute (EX), memory (MEM), write back (WB),
I-cache, D-cache, Integer register file, Floating register file and shared
FPU are existing components but slightly modified in order to support the ISA
of the Microgrid. The scheduler, allocator, network, thread table and family
table are the newly added components to support the concurrency constructs of
the microthreading model. The thread table and family table are used to store
the thread contexts and family contexts. Every core also has a single exclusive
context for creating an exclusive family and thread. The shared channels in the
microthreading model are implemented in the registers of the cores. These
registers are synchronizing and provide two ports for synchronous and
asynchronous completion e.g. D-cache and FPU operations are completed
asynchronously and therefore connected to the asynchronous ports of the
registers. The bus interface connects the L1-cache to the snoopy bus and the
network interface connects the core to the network of other cores on the
chip.

The threads in hardware are supported by the thread management unit through the
instruction set of the core for creation and synchronization of threads. It has
lowered the overhead of creation and synchronization of software threads from
more than ten thousands of cycles to just few cycles. It also has extremely
low overhead of context switching (zero cycle) and provides fine-grained
interleaving i.e. interleaving at every cycle. However, this interleaving is
not as strict as in traditional threads. Threads can execute multiple
instructions from a single thread until a context switch is required.
Interleaving can be bypassed to maximize pipeline usage if there is no other
available thread. In single-threaded programs interleaving does not make sense
anyway. To ensure fairness in the execution of threads, interleaving is
enforced so that no thread can monopolize the execution time of the core.

The binary code of a program generated for the microthreaded core can be
executed on any other core that supports the same instruction set as Microgrid.
The other core may not support concurrent execution and the code will be
executed as single threaded. The code can also be executed on different number
of thread slots per core. This is called binary code compatibility i.e. the
binary code may not get the same performance but at least it will not fail, its
performance will be gracefully degraded.

\begin{figure}

\begin{centering}

\includegraphics[width=0.8\textwidth]{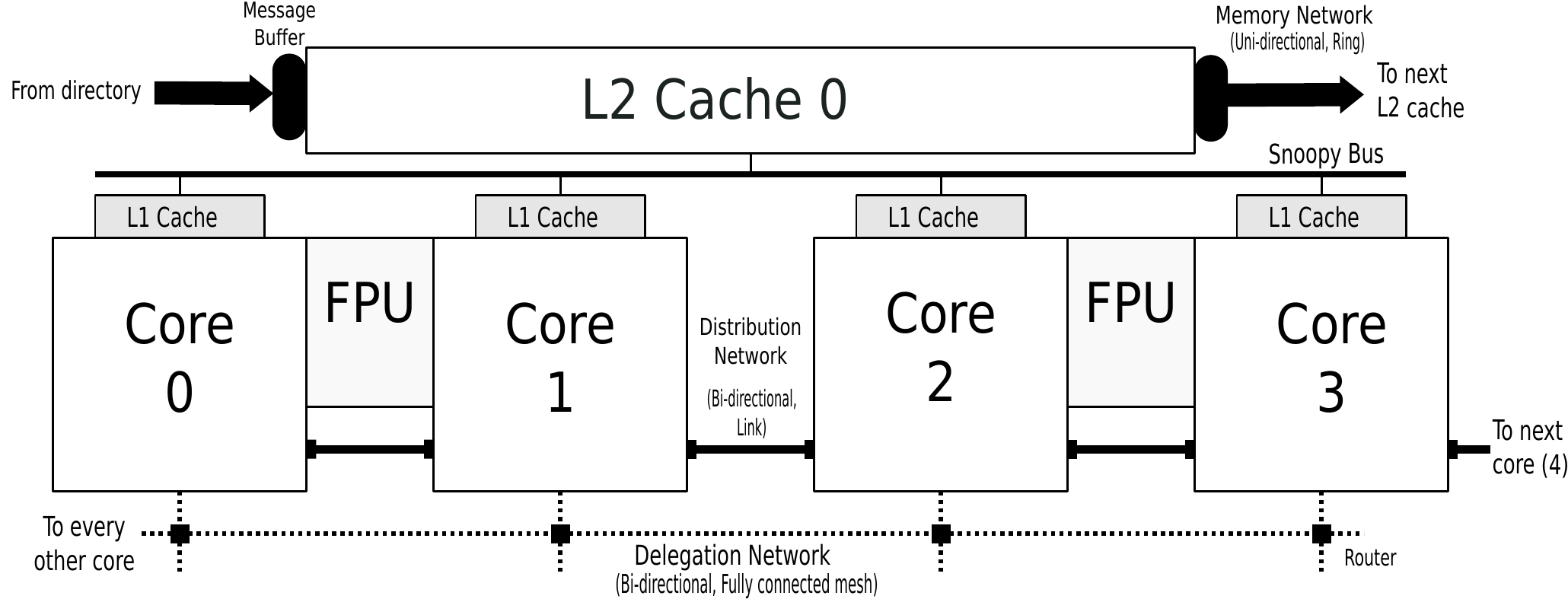}

\caption{\label{fig:multi_cores}A group of four cores, each core has an
L1-cache, two cores share an FPU and four cores share an L2-cache.}

\end{centering}

\end{figure}

\begin{figure}

\begin{centering}

\includegraphics[width=0.9\textwidth]{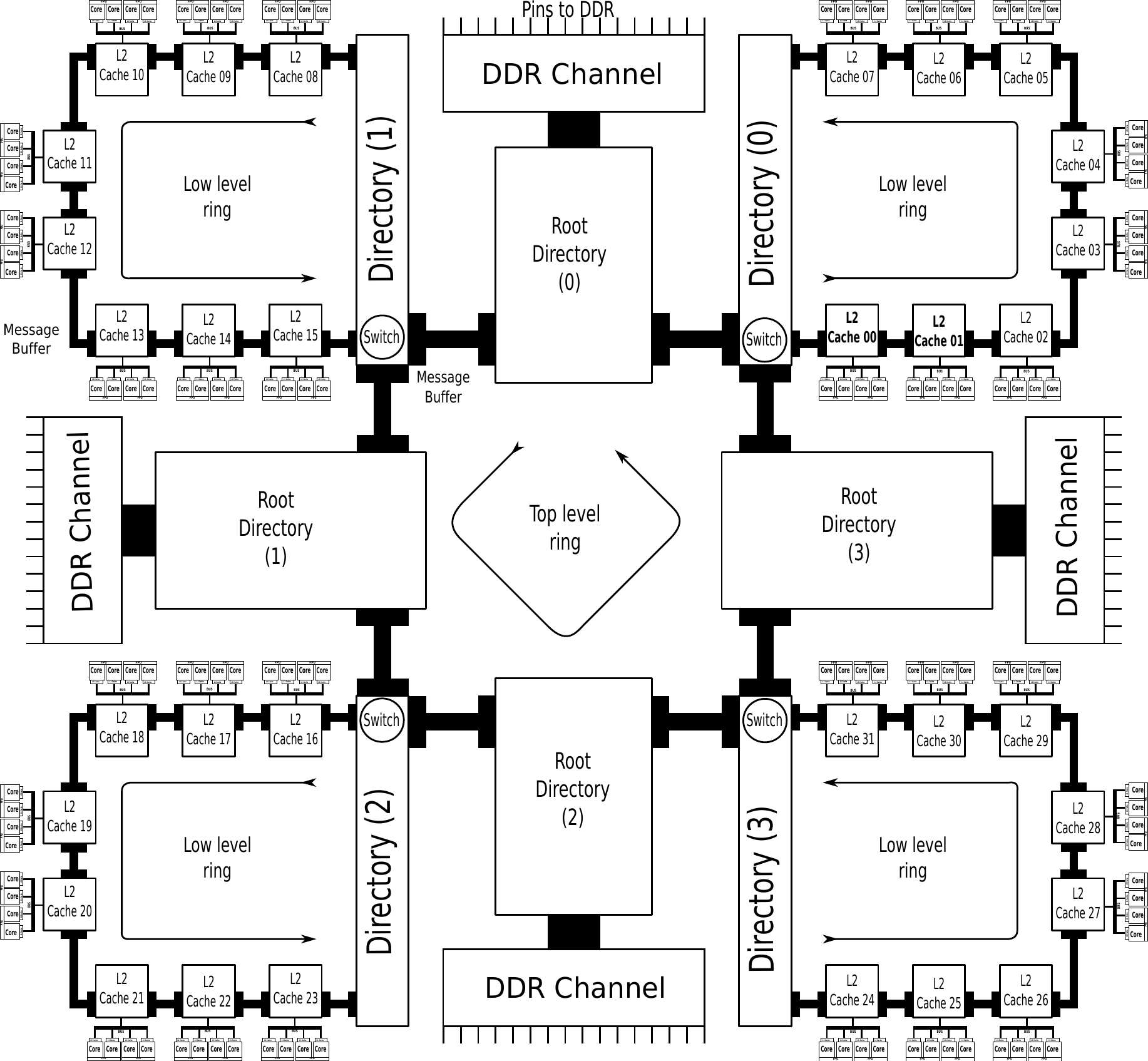}
    
\caption{\label{fig:microgrids_chip}The layout of 128 cores on the Microgrid
chip.}

\end{centering}

\end{figure}

As Microgrid provides many cores on a single chip, we show a group of four
cores in fig.~\ref{fig:multi_cores} to demonstrate the interconnection of large
number of cores. The delegation network is a Network-on-Chip (NoC) where all
cores are addressable from all other cores. Currently it is implemented as a
fully connected network in MGSim, but could be implemented as a mesh network in
the hardware implementation. A core is connected to the previous and the next
core by a distribution network. Two cores share an FPU, and every core has an
L1-cache which are connected to L2-cache by a snoopy bus. L2-caches are
connected with each other in the distributed cache network. In order to handle
deadlock in memory network, every L2-cache has an incoming and outgoing
message buffer. An example layout of the 128-core Microgrid in a
single chip is shown in fig.~\ref{fig:microgrids_chip}.

\subsection{The communication network}\label{com_network}

The Microgrid chip has two communication networks on the chip, in addition to
the distributed memory network. 

\subsubsection*{Delegation network}

Every core is connected to every other core by a bi-directional fully connected
network. This network is highly efficient because a message travels in 10 or
even less cycles from source to destination. The current implementation of the
Microgrid assumes a single cycle, node-to-node routing in a lightly loaded
network, but in a loaded network it may take more cycles due to contention and
buffering delays. Since delegation network is very efficient, it is used only
for the concurrency management between cores that are not adjacent.

\subsubsection*{Distribution network}

All the cores are connected in a single bi-directional daisy-chained network in
a Moore curve. The curve can be chosen in a way to preserve locality in
L1-caches and L2-caches. It takes two cycles to travel from one core to an
adjacent core. The cycles taken by a return trip of a message from a core to
any other core on the distribution network is given
in Eq:~\ref{eq:delay_distribution}, where $c$ is the number of cores. 

\begin{equation}
\label{eq:delay_distribution}
delay = 2 \times 2 \times c
\end{equation}

The distribution network is used for the logical partitioning of the chip i.e.
different parts of the program can be executed on different parts of the chip.
It provides grouping of cores to be used for delegating a family, where the
distribution network knows the starting core in the group and the size of the
group.

\subsection{Resource management in the hardware of the Microgrid}

An integer value in the source program is used to identify a group of adjacent
cores where a family can be delegated. We refer to the group of cores as
\emph{place} and the integer value that identify the place as \emph{placeid}.
The identifier can identify the starting core of the place and the size of the
place. More details about places can be found in~\cite{place_mgsim,poss.12.sl}
and~\cite[Chap.~11 \& App.~E]{poss.12}.

A family can execute on the same core as the parent thread or the same place as
the parent thread. The former is called \emph{local} place and the later is
called \emph{default} place and are identified by place id as $0$ and $1$
respectively. The execution of a family on a place different than used by the
parent thread is called \emph{delegation}. We can perform some arithmetic
operations on \emph{placeid} to derive the starting core and the size in the
place. The starting core can be calculated as given in
Eq:~\ref{eq:starting_core} and the size of the place can be calculated as given
in Eq:~\ref{eq:size}. The size of the place is always in power of 2, and the
starting core must have the \emph{coreid} which is multiple of the size of
the place. This is a restriction, but easy in implementation as this
information can be derived from integers.

\begin{equation}
\label{eq:starting_core}
starting\_core \;\;=\;\; (placeid\;\; \&\;\; (placeid\; -\; 1))\; >>\; 1
\end{equation}

\begin{equation}
\label{eq:size}
size \;\;=\;\; placeid\;\; \&\;\; -placeid
\end{equation}

\subsection{Proposed software service to access resources in the
Microgrid}

A software layer for the allocation of resources is defined in a protocol
referred as SEP~\cite{SEP,sl16} which provides an easy to implement and
efficient management of resources. In the initial research work, a single core
on the Microgrid is reserved for the operations of SEP. It works in mutually
exclusive manner, and can soon become a bottleneck when a lot of requests are
coming to this single core. Some research is ongoing in avoiding bottleneck and
making the management hierarchical.

\begin{figure}

\begin{centering}

\includegraphics[width=1\textwidth]{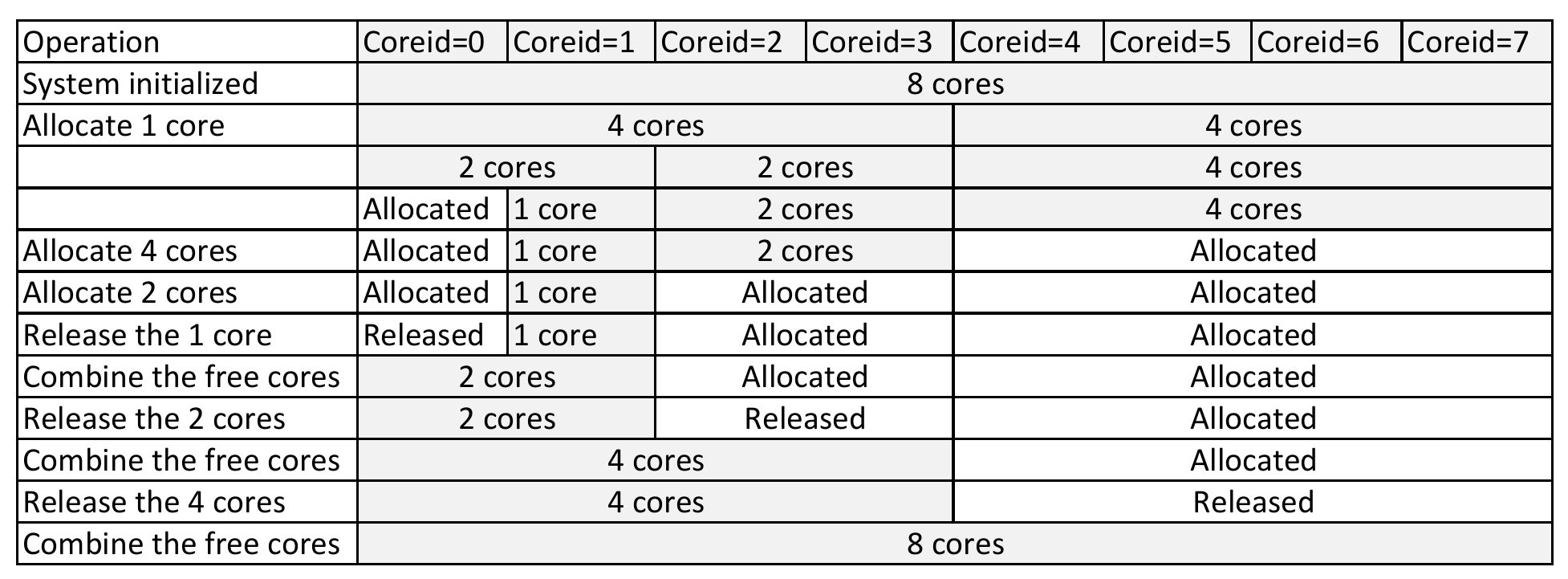}

\caption{\label{fig:sep_buddy_allocation}An example of the binary buddy
allocation of cores by SEP.}

\end{centering}

\end{figure}

SEP is implemented using the well-known binary buddy
allocation~\cite{Peterson:1977:BS:359605.359626} of the memory management. An
example of buddy allocation used for cores in the Microgrid is shown
in fig.~\ref{fig:sep_buddy_allocation}. It shows a system having a total of eight
cores with their identifiers given on the top. The time the system is
initialized, all the cores are in one group. Then a request is issued to
allocate one core and involves these steps:

\begin{itemize}

\item Divide 8 cores in half.

\item Still bigger than required, divide 4 cores into half.

\item Still bigger than required, divide 2 cores into half.

\item We found a single core as requested, allocate the single core.

\end{itemize}

A request for a given number of cores, will allocate the group of cores if
available. In case the available group is larger than requested, divide the
bigger group into smaller group until the group of requested cores is obtained.
When the allocated cores are released, they are grouped into a larger group. In
the given example when the last 4 cores are released it is grouped into larger
group of 8 cores, bringing the system to the initial state.

A programmer can choose different numbers of cores in a place. Every time a
place is requested, a group of cores which are not allocated to any other
family is allocated. The allocated cores can be used by the family (and sub
families if any) to which the place is allocated, and can not be allocated to
any other family until explicitly de-allocated. SEP provides two API to
programmers; $sep\_alloc()$ and $sep\_free()$ for allocation and de-allocation
respectively similar to $malloc$ and $free$ in C. The allocation and
de-allocation of SEP can be performed asynchronously, but in the current
implementation these processes are not asynchronous. The moment the allocation
or de-allocation is issued by a thread, the thread will wait until the
operation is completed. The allocation of place depends on the policy given
below:

\begin{itemize}

\item Minimum: Allocate at least the number of cores specified (may be more if
    available).

\item Maximum: Allocate at most the number of cores specified (may be less if
    unavailable).

\item Exact: Allocate the exact number of cores. The allocation will fail if
    the given number of cores can not be allocated.

\item Any size: Any available size in the power of 2 will be allocated,
    starting from the lowest available size of the group.

\end{itemize}

\section{Programming the Microgrid}\label{sn:program_microgrids}

In this section we show an example program i.e. Matrix Multiplication of equal
sized matrices, to demonstrate the way programs are written for the Microgrid.
The objective is to show that a sequential C program can be transformed to a
microthreaded program easily by high-level programming languages, a compiler or
with little effort by the programmer. We also show the concurrency constructs
in the generated assembly to demonstrate the way the concurrency constructs are
supported in the ISA of the Microgrid.

\subsubsection*{Sequential C program}

The sequential C code for Matrix Multiplication of size $1000 \times 1000$ is
shown in listing~\ref{code:seq_matmul}. We allocate three arrays; two for
source matrices and one for the result matrix. Once the memory is allocated, we
can fill these arrays with some numbers, but for saving space in the page, we
assume the existing values present in those memory locations. To perform matrix
multiplication we write three loops; outer loop, middle loop and inner loop.
The inner loop is the one which multiplies the two elements of the source
matrices and stores in the result matrix. After the multiplication is
completed, we free the allocated memory.

\begin{figure}
\begin{minipage}{\textwidth}
\scriptsize
\lstset{float, language=C, caption=Sequential execution of Matrix Multiplication.,
label=code:seq_matmul} 
\begin{lstlisting}
#include <stdio.h>

int N = 1000;

int main()
{
    // Allocate memory for matrices
    int *A = (int*) malloc(N * N * sizeof(int));
    int *B = (int*) malloc(N * N * sizeof(int));
    int *C = (int*) malloc(N * N * sizeof(int));

    int i, j, k;

    // Perform the multiplication
    // matmul_outer
    for(i=0; i<N; i++)
    {
        // matmul_middle
        for(j=0; j<N; j++)
        {
            //matmul_inner
            for(k=0; k<N; k++)
            {
                C[i*N+j] += A[i*N+k] * B[k*N+j];
            }
        }
    }

    free(A), free(B), free(C);
    return 0;
}
\end{lstlisting}
\end{minipage}
\end{figure}

\subsubsection*{Microthreaded program}

The microthreaded program for the Matrix Multiplication of size $1000 \times
1000$ written in SL is shown in listing~\ref{code:par_matmul}. In the $t\_main$
function, we first allocate a group of 8 cores in the Microgrid. Then we
allocate three arrays for source and result matrices. Then we replace the outer
loop by creating a family of $N$ threads. The threads in the outer family
create further middle families and then each of those creates the inner family.
The inner family perform the multiplication of elements in source arrays and
store the multiplication in the resultant array. The transformation from
sequential program to microthreaded program involves the creation of families
and shared and global parameters. Once all the families are synchronized, the
allocated memory to arrays and allocated cores are released.

\begin{figure}
\begin{minipage}{\textwidth}
\scriptsize
\lstset{float, language=C, caption=The microthreaded version of Matrix Multiplication.,
label=code:par_matmul} 
\begin{lstlisting}
#include <svp/sep.h>
#include <stdio.h>
int N = 1000;

//matmul_inner
sl_def(matmul_inner, void,
    sl_shparm(long, sum), sl_glparm(void*, A),
    sl_glparm(void*, B), sl_glparm(size_t, i),
    sl_glparm(size_t, j)) {
    sl_index(k);
    int (*A)[N][N] = (int (*)[N][N])(void*) sl_getp(A);
    int (*B)[N][N] = (int (*)[N][N])(void*) sl_getp(B);

    int v = (*A)[sl_getp(i)][k] * (*B)[k][sl_getp(j)];
    sl_setp(sum, v + sl_getp(sum));
}sl_enddef

//matmul_middle
sl_def(matmul_middle, void,
    sl_glparm(void*, A), sl_glparm(void*, B),
    sl_glparm(void*, C), sl_glparm(size_t, i)) {
    sl_index(j);
    sl_create(,,0,N,,,,matmul_inner,
        sl_sharg(long, sum, 0), sl_glarg(void*, , sl_getp(A)),
        sl_glarg(void*, , sl_getp(B)), sl_glarg(size_t, , sl_getp(i)),
        sl_glarg(size_t, , j));
    sl_sync();

    int (*C)[N][N] = (int (*)[N][N])(void*) sl_getp(C);
    (*C)[sl_getp(i)][j] = sl_geta(sum);
}sl_enddef

//matmul_outer
sl_def(matmul_outer, void,
    sl_glparm(void*, A), sl_glparm(void*, B),
    sl_glparm(void*, C)) {
    sl_index(i);
    sl_create(,,0,N,,,,matmul_middle,
        sl_glarg(void*, , sl_getp(A)), sl_glarg(void*, , sl_getp(B)),
        sl_glarg(void*, , sl_getp(C)), sl_glarg(size_t, , i));
    sl_sync();
}sl_enddef

sl_def(t_main, void) {
    // Allocate a place using SEP
    int core = 8;
    sl_place_t pid;
    if (sep_alloc(root_sep, &pid, SAL_EXACT, core) == -1) {
        printf("cannot allocate a place.");
        exit(1);
    }
    
    // Allocate memory for matrices
    int *A = (int*) malloc(N * N * sizeof(int));
    int *B = (int*) malloc(N * N * sizeof(int));
    int *C = (int*) malloc(N * N * sizeof(int));

    // Perform the multiplication
    sl_create(,pid,0,N,,,,matmul_outer,
        sl_glarg(void*, , A), sl_glarg(void*, , B), sl_glarg(void*, , C) );
    sl_sync();

    free(A); free(B); free(C); sep_free(root_sep, &pid);
} sl_enddef
\end{lstlisting}
\end{minipage}
\end{figure}

\subsubsection*{Microthreaded assembly code}

The microthreaded assembly code is shown in listing~\ref{code:assembly_matmul}. We
can see the assembly instructions of concurrency constructs i.e. $allocate$ for
the allocation of family, $crei$ for creation of family, $sync$ for
synchronization of family, $puts$ writing a shared variable, $gets$ reading
from a shared variable etc. This is the innovation in the Microgrid, which
avoids the mapping of software threads to hardware threads, which reduces the
overhead from more than 10-100 thousands cycles of software threads to just few
cycles in hardware threads. The ISA of
program is scheduled by the microthreaded core with the hardware support for
concurrency management.

\begin{figure}
\begin{minipage}{\textwidth}
\scriptsize
\lstset{float, language=C, caption=A snipper of the assembly program generated from the
microthreaded program of Matrix Multiplication for the Microgrid., label=code:assembly_matmul}
\begin{lstlisting}
...
# Assembly code of thread matmul_inner
.ent matmul_inner
.registers 4 1 4 0 0 0
matmul_inner:
ldpc $l3
ldah $l3, 0($l3)
lda $l3, 0($l3)
$matmul_inner..ng:
ldah $l1,N($l3)		
ldl $l1,N($l1)		
mulq $l0,$l1,$l2
mulq $g2,$l1,$l1
addq $l2,$g3,$l2
addq $l1,$l0,$l0
s4addq $l2,$g1,$l2
s4addq $l0,$g0,$l0
ldl $l2,0($l2)
ldl $l0,0($l0)
mull $l2,$l0,$l0
addq $l0,$d0,$l0
mov $l0, $s0
end
.end matmul_inner
...

# to show the concurrency constructs in the ISA
.ent t_main
...
allocate/s $l0, 0, $l0
setstart $l0, 0
ldq $l1,matmul_outer($l17)
setlimit $l0, $l9
setstep $l0, 1
setblock $l0, 0
wmb
crei $l0, 0($l1)
putg $l12, $l0, 0
putg $l11, $l0, 1
putg $l10, $l0, 2
sync $l0, $l1
mov $l1, $31
release $l0
...
.end t_main
\end{lstlisting}
\end{minipage}
\end{figure}

\subsubsection*{SL tool-chain}

\begin{figure}

\begin{centering}

\includegraphics[width=0.6\textwidth]{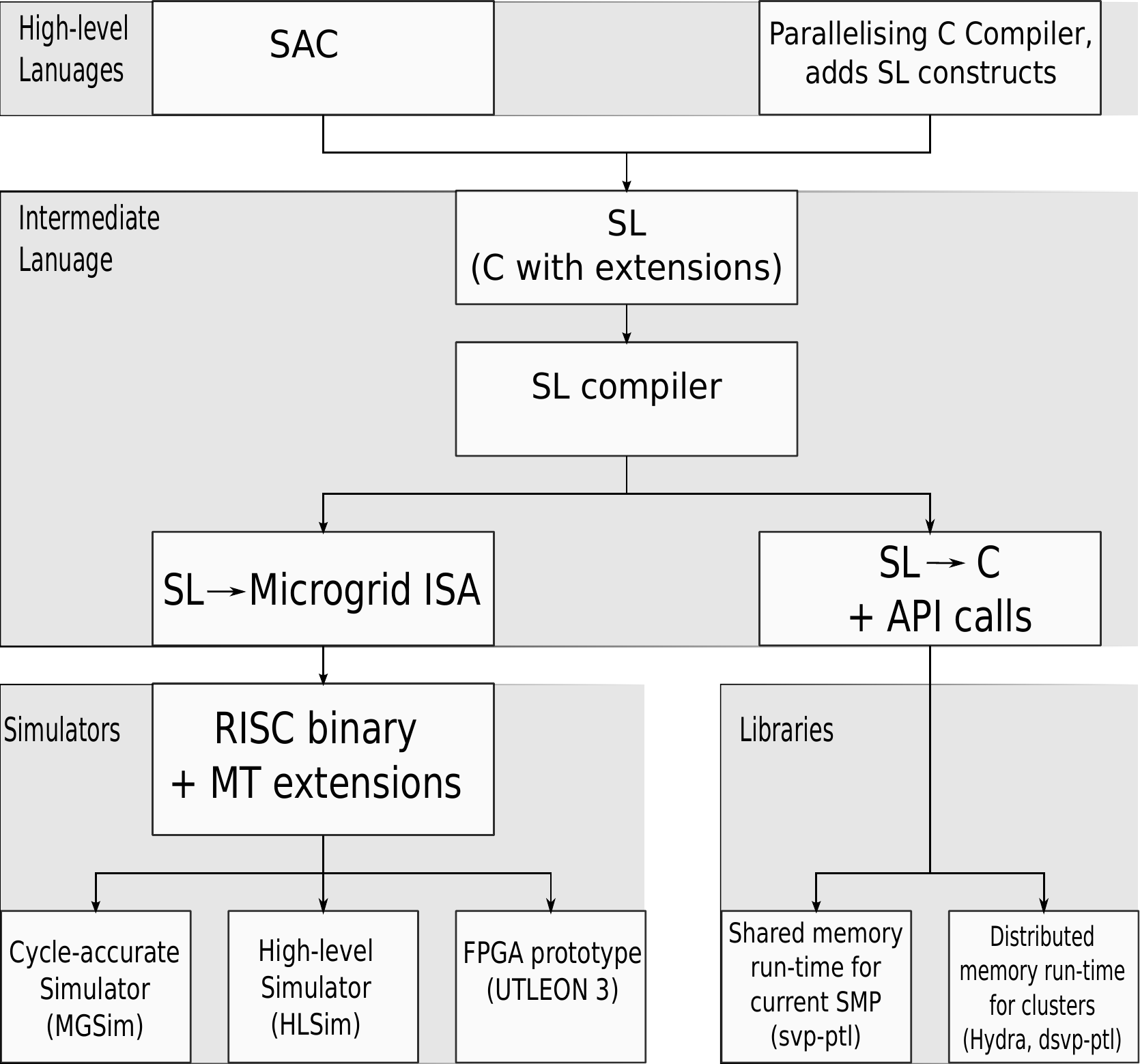}

\caption{\label{fig:svp_toolchain}The SL tool-chain.}

\end{centering}

\end{figure}

The SL tool-chain is shown in fig.~\ref{fig:svp_toolchain} and is built around
SL~\cite{poss.12.sl} which is a C based language but is extended to express the
concurrency constructs of the microthreading model. SL is an intermediate
language intended for higher level programming languages (such as Single
Assignment C or
SAC~\cite{sac_tutorial:,grelck.09.cpc,GrelckJFP05,GrelSchoIJPP06,
Grelck:2007:SOS:1248648.1248654},
FastFlow~\cite{Aldinucci12fastflow:high-level,ff:distr:cgs:12} etc.) and
parallelizing C compiler~\cite{saougkos.09.cpc,saougkos.11}. SAC is Matlab-like
programming language, and it provides concurrency from a very high view.
Programmer writes program using arrays and let the compiler decide to exploit
concurrency in the programs.

A number of tools and simulators are added to the designer's toolbox and used
for the evaluation of the Microgrid from different perspective. The SL compiler
can generate binary for different implementations of the Microgrid. We have
software libraries that provide the run-time systems for the microthreading
model on the shared memory SMP machines and referred as
\emph{svp-ptl}~\cite{SVP-PTL2009} and distributed memory for clusters/grids and
are referred as Hydra~\cite{Andrei:msc_hydra:2010} and
\emph{dsvp-ptl}~\cite{DSVP-PTL2011} The SL compiler can generate binary for
UTLEON3~\cite{5491777,danek.12},
MGSim~\cite{Bousias:2009:IEM:1517865.1518255,poss.13.MGSim.SAMOS} and
HLSim~\cite{Irfan:multipe_levels_hlsim:2013, Irfan:oneipc_hlsim:2013,
Irfan.12.2013.signatures, Irfan.12.2013.CacheBased, Irfan.01.2014.analytical,
Irfan:hl_sim_ptl:2011, Irfan:msc_hlsim:2009, Uddin:2012:CSM:2162131.2162132}.

Unless specified otherwise, the SL compiler generates two implementations for
every family in the given SL program; concurrent and sequential. Either one is
used as per the dynamic state of the chip. In case resources can not be
allocated to a family, the sequential version can be used. The sequential
version of the program is used to avoid deadlock and if the deadlock can be
avoided by static analysis, then suspension on resource allocation is safe. The
programmer can force the program to use the concurrent version all the time
i.e. threads are suspended until resources become available. 

\section{The microthreading model in the context of the Microgrid}
\label{sn:microthreading_microgrid}

We gave details of the Microgrid, and would like to revisit the microthreading
model in the context of the Microgrid. We need to understand all the details of
the model and the architecture in order to simulate the architecture at a
high level, as the high-level simulator have to exhibit the same behavior.

\subsection{A concrete example}

\begin{figure}

\begin{centering}

\includegraphics[width=0.4\textwidth]{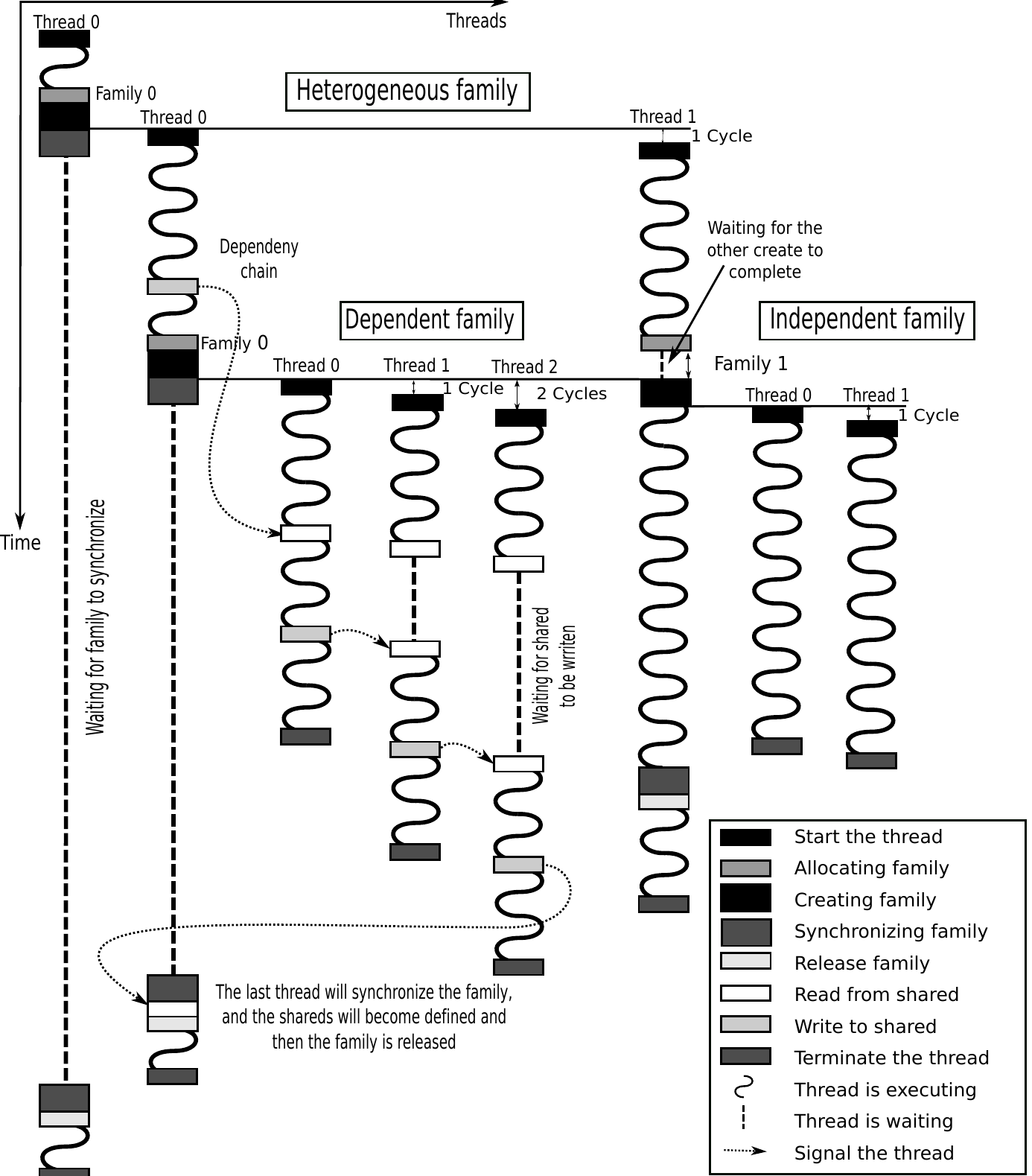}

\caption{\label{fig:concurrency_tree}A concrete example of the microthreaded
program of the microthreading model executing in the single core of the
Microgrid. We explicitly use a single core to show the coordination of
different events in different threads over time.}

\end{centering}

\end{figure}

A concrete example of the microthreaded program in the microthreading model
executing in a single core of the Microgrid is shown
in fig.~\ref{fig:concurrency_tree}. All concurrency constructs are shown in the
figure in the form of rectangles of different sizes/color. We also show the
effect of concurrency constructs on the execution of threads. Every thread has
an implicit \emph{start} and \emph{end} event. Threads are created at the rate
of one thread per cycle. The creation process takes 4 cycles (with additional
latency to load the cache-line from memory if not in the I-cache, also
potential queueing delays if another family is being created). The instructions
of a thread can be executed as soon as the thread is created which means the
first instruction of the first thread can execute along with the creation of
the other threads~\cite{mike.12}. The process of creation of threads is
implemented sequential on a single core, and is decided based on the trade-off
between simple design and efficient creation of family. In the figure we can
see that two \emph{allocate} events execute concurrently, but the \emph{create}
events are sequentialized on a single core.

We show a heterogeneous family of threads i.e. Threads $0$ and $1$ in Family
$0$ in the first level is different from each other. We show a dependent family
where threads are waiting for the previous threads to write the shared
variable/register. We also show that the sync event is waiting for all the
threads to complete. We show a homogeneous independent family where the parent
threads is executing along with the child threads and therefore the sync does
not wait as the time the sync is issued where all the created threads are
completed already.

\subsection{Family's Life cycle}\label{sn:family_lifecycle}

\begin{figure}

\begin{centering}

\includegraphics[width=1\textwidth]{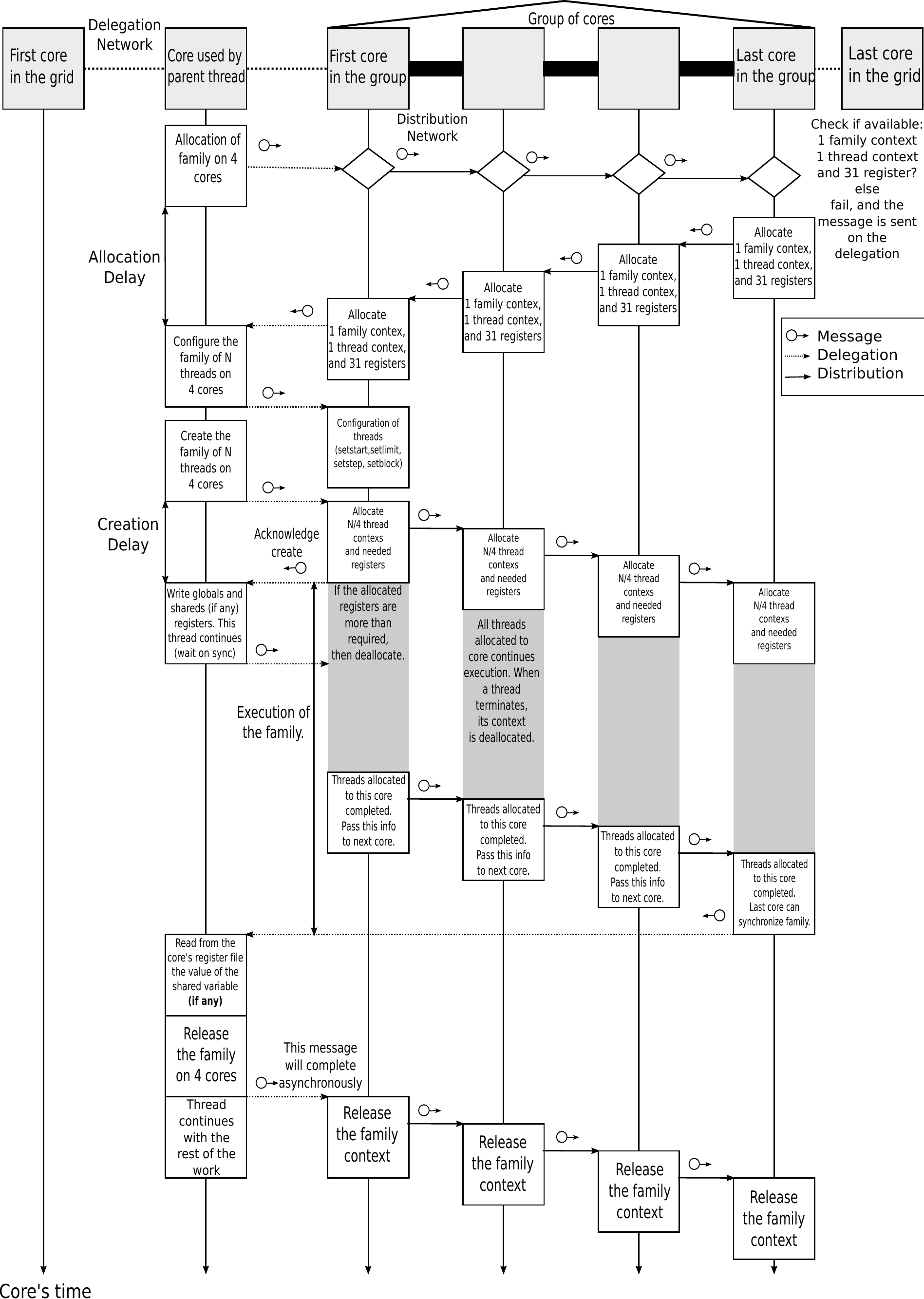}

\caption{\label{fig:family_lifecycle}The life cycle of a family during the
execution in the Microgrid.}

\end{centering}

\end{figure}

A family passes through different stages in the Microgrid during the execution.
In this section we consider an example when a parent thread is executing on a
core and it delegates a family to a place of four cores shown
in fig.~\ref{fig:family_lifecycle}. A detailed explanation of these stages is
given in below section. 

\subsubsection*{Allocation}\label{sn:allocation_families}

It is implemented as try-to-allocate mechanism (see below for various options
on allocate). In this process; family context, thread contexts and registers on
all cores of the place are tried to be allocated. The parent thread sends a
message on the delegation network to the first core. The first core checks
the availability of at least one thread context, one family context and 31
registers. In case of success the message is passed to the next core of the
place using distribution network. When all the cores succeed, the message comes
back from the last to the first core on distribution network, and the contexts
are asynchronously allocated. The first core then notifies the parent core
using delegation network. The allocate protocol has three different modes:

\begin{itemize}

\item Normal (or default): The allocation fails immediately after the
    allocation fails on any core in the place.

\item Suspend: The allocation waits for the availability of resources.

\item Exclusive: The allocation allocates exclusive contexts and in case of
    unavailability, keeps waiting until resources are acquired.

\end{itemize}

The allocate protocol has four different strategies:

\begin{itemize}

\item Normal (or default): It will try to allocate as many cores as requested,
    possibly down to one core (in power of 2).

\item Exact: It will allocate the family on all the requested cores.

\item Single: It will allocate the entire family to the first core in the
    place.

\item Balanced: It will allocate the entire family to the least loaded core of
    the place, in terms of family contexts.

\end{itemize}

\subsubsection*{Configuration}

Once the contexts are allocated on the cores, the next message is sent on the
delegation network to the first core to store parameters in the allocated
family table entry for the bulk creation. It consists of $setstart$,
$setlimit$, $setstop$, $setblock$ etc. which are derived from the create
construct and defines the number of threads to be created per core. The
configuration process is completed asynchronously and does not need any
acknowledgement. Therefore as soon as the configuration message is issued,
the next message can also be sent. In the current implementation, the
delegation network preserves the order of messages i.e. configure messages will
reach to the destination before create message.

\subsubsection*{Creation}

The create message is sent from the parent core to the first core in the place
via delegation network. As soon as the message is received by a core, thread
creation is started asynchronously and the message is forwarded to the next
core in the place. The first core will acknowledge the parent core when all the
threads on the first core are created. The parent core may then write the
global and/or shared registers to the first core through inter-context
communication and can continue with its instructions (if any) or wait for the
synchronization of the family.

When all the threads are created, the core can de-allocate the contexts that
are not really required e.g. at the time of allocation 31 registers were
allocated, but suppose a thread is created which only uses 15 registers
therefore the core will de-allocate the extra 16 registers. All the threads
continue their execution on the allocated core, and as the threads get
terminated the context get de-allocated asynchronously. Every core shares the
information when it completes executing the allocated threads of a given family
with the next core. The last core has the information that all the threads in
the family are completed.

\subsubsection*{Synchronization}

When all threads in the family complete, the last core will send a message on
the delegation network to the parent core. The parent thread is activated so
that the family can be synchronized. All the modified state from the family
becomes defined. In the case of a dependent family, the parent core will also
read the value from the register updated by the last thread through
inter-context communication.

\subsubsection*{Release}

A message is sent from the parent core to the first core on delegation network
to release the family. This process will complete asynchronously and the parent
core can continue its execution.

\subsection{Thread's life cycle}\label{sn:thread_lifecycle}

\begin{figure}
\begin{centering}
\label{Neighborhood}
    \includegraphics[width=0.8\textwidth]{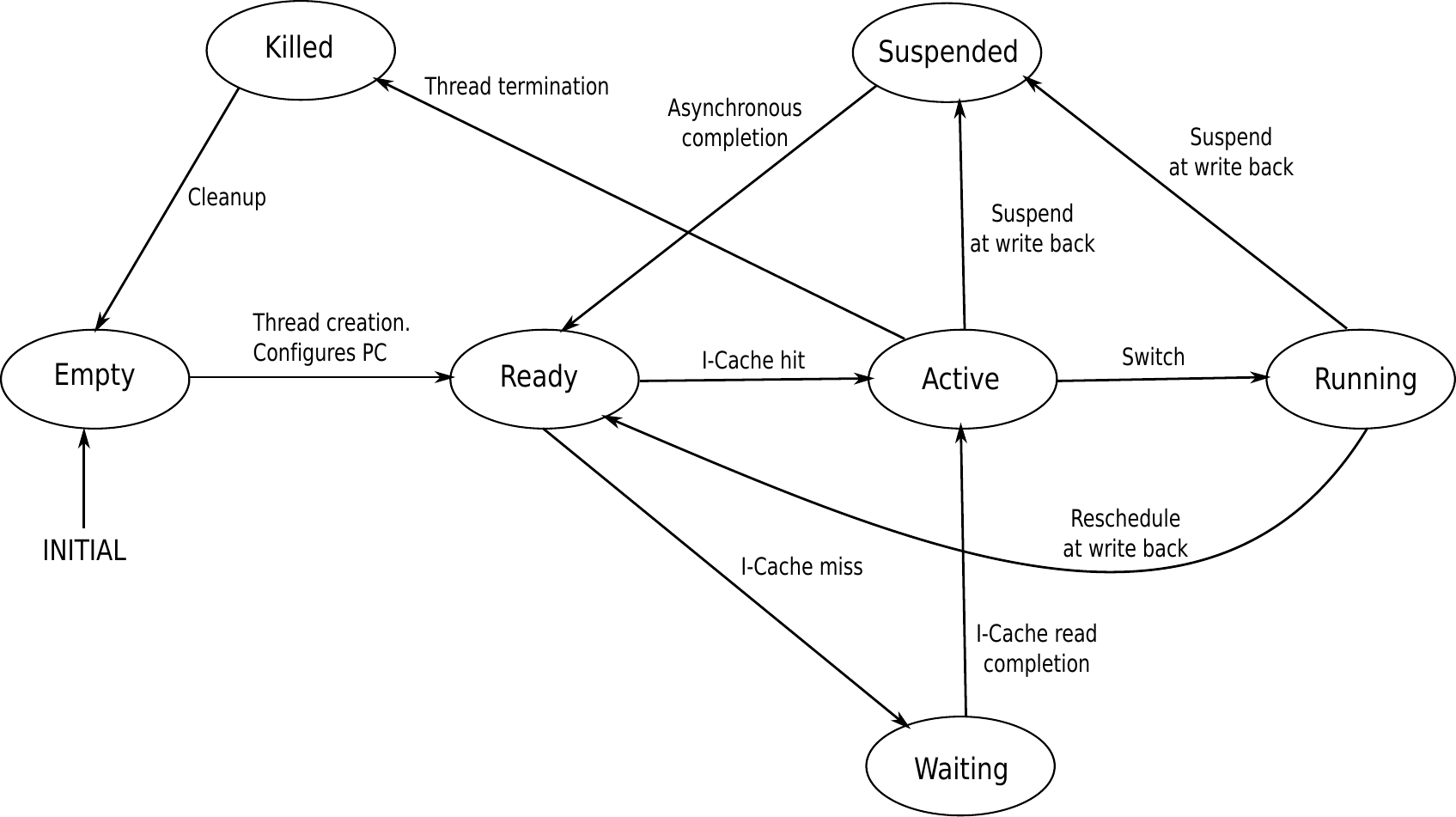}
    \caption{\label{fig:thread_lifecycle}The lifecycle of a thread during the execution on the core of the
    Microgrid.}
\end{centering}
\end{figure}

A thread passes through different stages during the execution and is shown
in fig.~\ref{fig:thread_lifecycle}. A newly created thread is allocated thread
context, registers and PC (Program counter) is configured to place the thread
in ready queue. In case the thread gets an I-cache miss the thread waits
until the instructions are loaded from cache. The thread then passes to the
active queue. The context of the thread is moved from the active queue to
running where the instructions actually execute in the execute stage of the
pipeline. A thread can be suspended when it is active or running at the write
back stage of the pipeline e.g. in case of dependency. The asynchronous
completion of a thread moves the thread from the suspended stage to ready
queue. At the termination of the thread the allocated entries are cleaned up
which can be used by any newly created threads.

\subsection{Distribution of threads}

\begin{figure}
\begin{centering}
\label{Neighborhood}
    \includegraphics[width=0.6\textwidth]{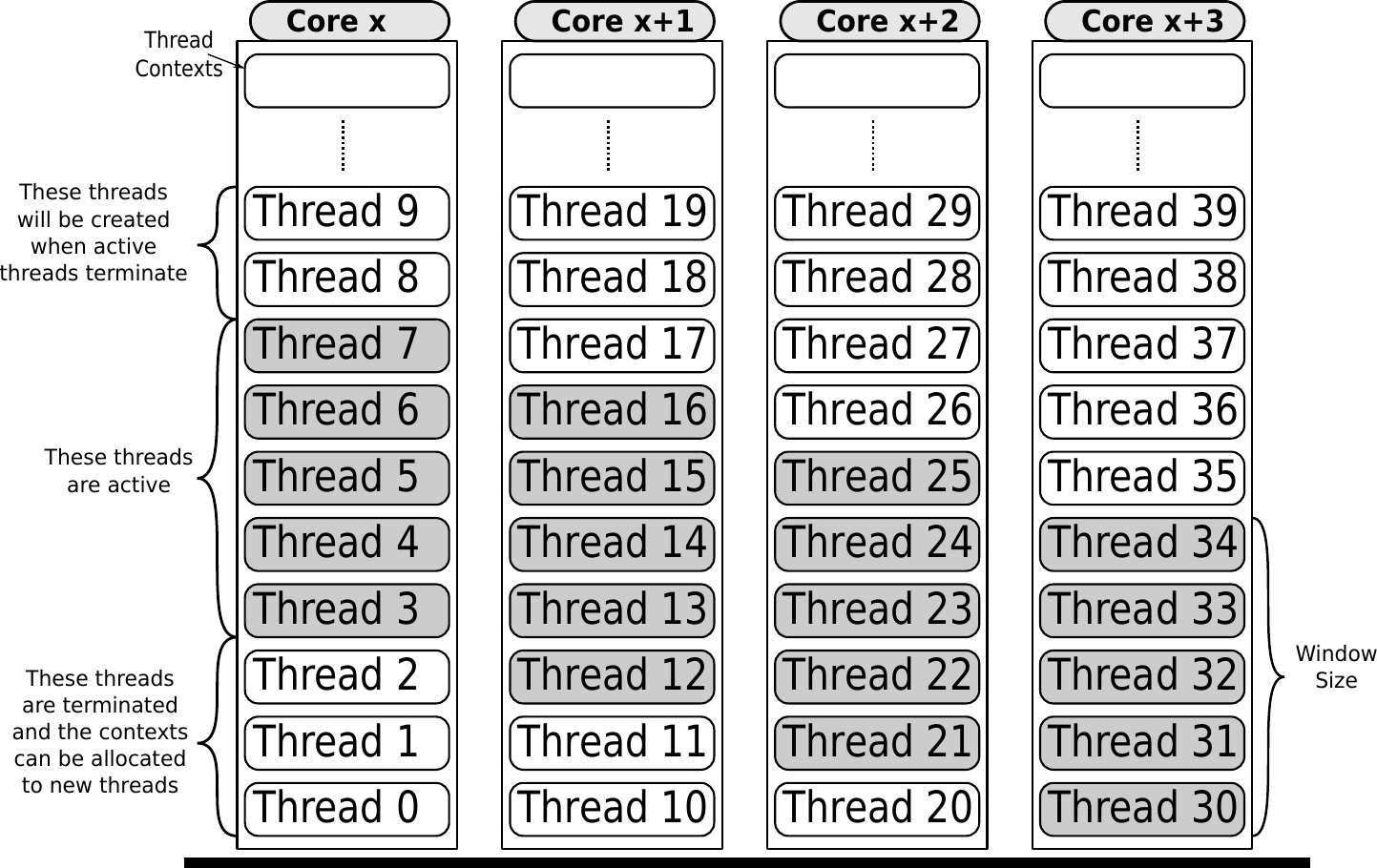}
    \caption{\label{fig:thread_distribution}Distribution of independent threads on 4 cores with 5 window size.}
\end{centering}
\end{figure}

The distribution of threads to the cores in a place depends on the type of
family i.e. dependent or independent. The threads in a dependent family are
restricted to single core only, as no significant speedup can be obtained by
distributing them on multiple cores. However instructions in the thread will
get benefit from latency tolerance on asynchronous completion in loads, store
and floating point operations.

Threads of an independent family are distributed by an equal distribution of
threads per core. In fig.~\ref{fig:thread_distribution} we show an example of a
family with 40 threads on 4 cores with window size of 5. Threads are
distributed as 40/4 and every core can create 10 threads. But the window size
is 5 and therefore only 5 threads can be activated at a time on a given core.
As soon as a thread terminates, a new thread can use the context of the
terminated thread. For instance on core x, threads 0,1 and 2 are terminated,
threads 3-7 are executing and threads 8 and 9 are still waiting to be
created.

\subsection{Communication through registers}\label{sn:registers}

\begin{figure}
\begin{centering}
    \includegraphics[width=0.6\textwidth]{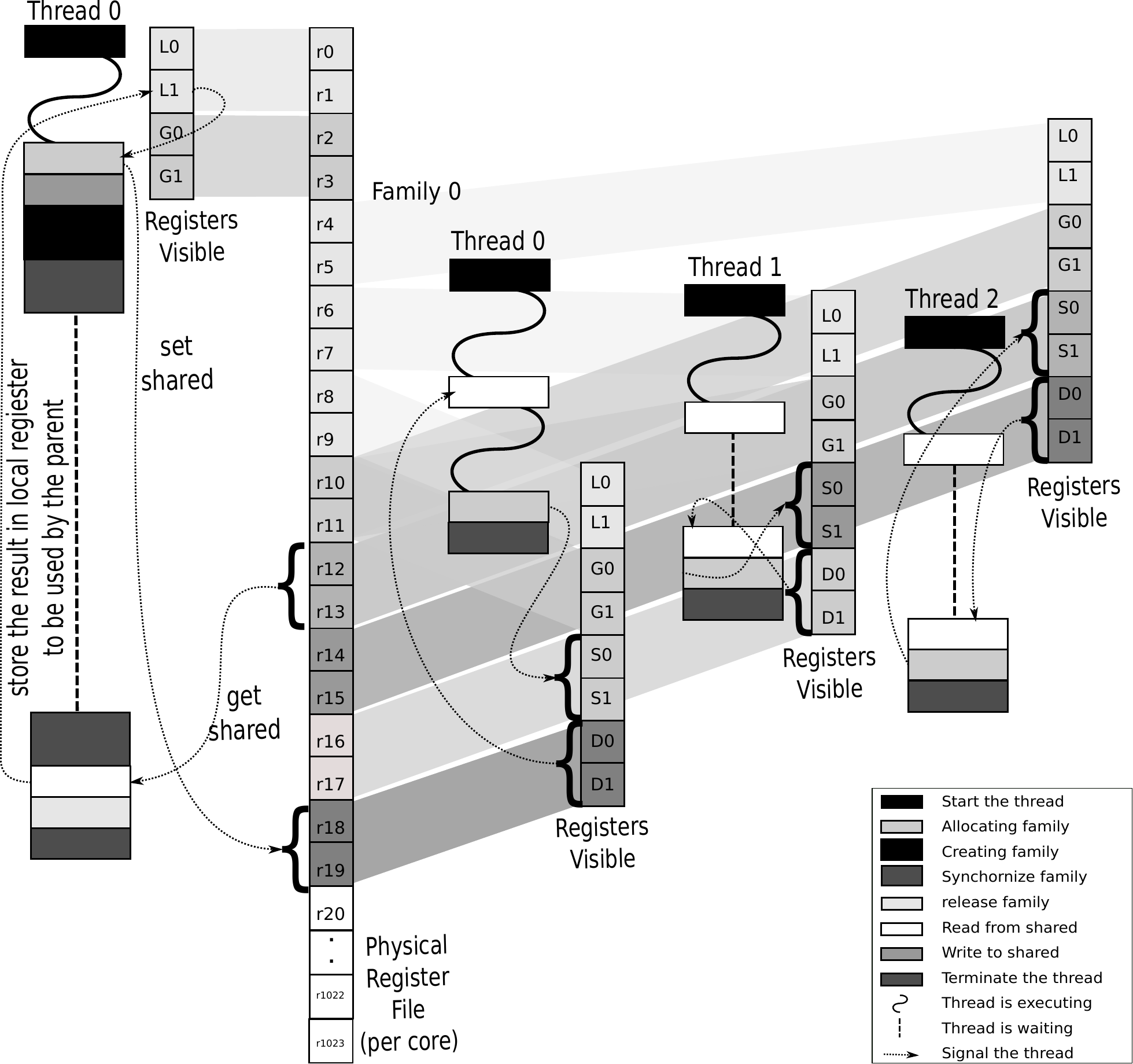}
    \caption{\label{fig:comm_registers}Allocation of registers to threads and their communication through
    registers.}
\end{centering}
\end{figure}

Every thread created in the Microgrid uses a set of registers in the register
file. The channels of the microthreading model used for the communication and
synchronization of the family introduced
in Section~\ref{sn:communitation_synchronization} are implemented in registers of the
Microgrid. The registers allocated to a thread are categorized as; globals,
locals, shareds and dependents. The mapping of registers to threads and their
communication through registers is shown in fig.~\ref{fig:comm_registers}. Some
explanation of these registers are given as:

\begin{itemize}

\item Global registers implements global channels and are visible to all
    threads in the family. The parent thread writes to the global registers and
    all created threads in the family can read from them.

\item Local registers are only visible to the individual thread only.

\item Dependent registers implements shared channels and have read-only access
    to the shared registers from the previous threads.

\item Shared registers implements shared channels and have the write access by
    the current threads, and carry the modified value to be read by the next
    thread.

\end{itemize}

\section{I/O in the Microgrid}
\label{sn:io_microgrid}

The Microgrid requires a decentralized approach where not every core is
connected to I/O but some specialized microthreaded cores are used to support
I/O and referred as I/O cores. I/O cores have limited/extended instruction
set~\cite{mgsim14} compared to the regular microthreaded core and has no
floating point operations. The instruction set of the I/O core supports the I/O
infrastructure and are connected to the delegation network of the Microgrid. We
do not simulate I/O cores in the current implementation of HLSim, therefore we
refer readers to~\cite{hicks09-io,poss.12,poss.12.rapido} for the I/O
management in the Microgrid.

\section{Conclusion}
\label{sn:conclusion}

The microthreading model is a hybrid dataflow model and provides the simplicity
of \emph{von Neumann} model and asynchronous completion of the \emph{dataflow}
execution model. It is an abstract machine model for many-cores architecture
and shifts the perspective from software threads to hardware threads. Because
of asynchronous completion and fine-grained latency tolerance the
microthreading model can potentially achieve the goal of RISC i.e. \emph{one
instruction per cycle} in the throughput of the program assuming single issue
width. The details of a particular type of future many-cores systems are given
to demonstrate the complexity of the architecture. 

\section*{Acknowledgement}
The author would like to thank Dr. Raphael Poss, Dr. Michiel van Tol and Prof.
dr. Chris Jesshope.





\bibliographystyle{plain}
\bibliography{main}







\end{document}